\documentclass[conference]{IEEEtran}
\IEEEoverridecommandlockouts

\usepackage{cite}
\usepackage{amsmath,amssymb,amsfonts}
\usepackage{algorithmic}
\usepackage{graphicx}
\usepackage{textcomp}
\usepackage{xcolor}
\usepackage{listings} 
\usepackage{subcaption}
\usepackage{listings-rust}% https://github.com/denki/listings-rust

\def\ioconsistency{input--output consistency}

\usepackage{listings}
\usepackage{ifthen}

\makeatletter
\let\MYcaption\@makecaption
\makeatother

\usepackage{caption}

\makeatletter
\let\@makecaption\MYcaption
\makeatother

\newcommand\sourcecodeposition{h}

\begin{document}

\title{RustSFQ: A Domain-Specific Language \\for SFQ Circuit Design}

\author{\IEEEauthorblockN{Mebuki Oishi, Sun Tanaka, Shinya Takamaeda-Yamazaki}
\IEEEauthorblockA{
\textit{The University of Tokyo}\\
\{mebuki, st, shinya\}@is.s.u-tokyo.ac.jp}
}

\maketitle

\begin{abstract}
Cell-based design of a single-flux-quantum (SFQ) digital circuit requires \textit{\ioconsistency{}}; every output signal must be consumed \textit{only once} by the input of the following component, which is a unique constraint, unlike the traditional CMOS digital circuit design.
While there are some cell libraries and simulation tools for SFQ circuit development, they do not verify the \ioconsistency{}, and designers have significant responsibilities to ensure it manually. 
Additionally, designers have to carefully manage net names without unintended duplication and correct connectivity among nets in a netlist for simulations.

We propose RustSFQ, a domain-specific language (DSL) embedded in Rust that automatically ensures the \ioconsistency{} in the SFQ circuit by leveraging the \textit{ownership system} of Rust.
Each SFQ circuit element is represented as a function while wires are represented as instances, and the Rust compiler verifies that multiple elements do not share a single wire through the ownership system.
Circuit descriptions in the RustSFQ are successfully compiled into low-level netlists for both analog and digital circuit simulations, and the DSL provides higher productivity than the conventional design flow.

Using the RustSFQ, we developed an SFQ-based Reed--Solomon encoder with a 4-bit width for the first time as a case study. We confirmed that the circuit operated correctly at 10~GHz through circuit simulations.

\end{abstract}


\maketitle

\section{Introduction}

Single-flux-quantum (SFQ) circuits are superconducting circuits operating at cryogenic temperatures. Due to their extremely high-speed operation and low power consumption, they attract significant interest as next-generation computing devices~\cite{likharev1991rsfq}.

SFQ circuits are driven by voltage \textit{pulses} unlike CMOS circuits, which are driven by voltage \textit{levels}. Each pulse corresponds to a single magnetic flux quantum, implying it does not spontaneously split or disappear. Thus, SFQ digital circuits must satisfy \textit{\ioconsistency{}}, a one-to-one correspondence between inputs and outputs. In other words, every output signal must be consumed \textit{only once} by the input of the following component, which is a unique constraint, unlike the traditional CMOS digital circuit design.

Another characteristic of SFQ circuits is that logic gates, such as AND or NOT, are clock-synchronous, as well as flip-flops. Consequently, while SFQ circuits can achieve extremely high clock frequency through deep gate-level pipelining, it requires careful consideration of how clock signals are supplied to each logic gate~\cite{ishida2020supernpu, takagi2014circuit}.

For higher productivity in SFQ circuit development, cell-based design~\cite{schindler2022coldflux} is preferred over full custom design using primitive circuit elements like resistors and inductors, as is done in CMOS circuit development. However, due to the pulse-driven nature, designers must ensure that the SFQ circuit satisfies \ioconsistency{}.

There is an analog circuit simulator~\cite{kirichenko2019josim} tailored to SFQ circuit evaluation, and such analog simulator requires a netlist of an SFQ circuit described in the SPICE format as the input. However, directly describing a netlist is a significant burden for designers because a unique \textit{net} name without unintended duplication should be assigned manually to each signal, and careful connectivity management is required.

To address these problems in SFQ circuit design, we propose a domain-specific language (DSL) enbedded in Rust.
Rather than other programming languages, we specifically use Rust to leverage its ownership system to facilitate static checking of \ioconsistency{}. Since designing a high-performance SFQ circuit requires fine-tuning of clock timings, this DSL is designed to describe a circuit at the gate level, close to the netlist. Moreover, the DSL contributes to efficient circuit design because it can generate netlists for both digital and analog simulations with proper net names automatically assigned.

As a case study using the RustSFQ, we developed a Reed--Solomon encoder. Reed--Solomon codes are a type of error correction code and have been applied in various fields such as QR codes and wireless communications, taking advantage of their high error correction capability~\cite{thach2014robust, daya2013fpga}.

\section{RustSFQ}
\subsection{Implementation and Basic Design}

RustSFQ is provided as a Rust crate, enabling circuit descriptions to be written as fully valid Rust code. 
The primary data types are \texttt{Wire} and \texttt{Circuit}. Methods on \texttt{Circuit} instantiate logic gates, and \texttt{Wire} is used to connect gates. Each \texttt{Wire} can only be used only once by leveraging Rust's ownership system.

The \texttt{Wire} struct internally holds only a single \texttt{String}. By wrapping it in a struct, constructor calls and cloning are restricted. The \texttt{Circuit} struct has a list of gate instances; when a gate function is invoked, a new gate instance is added to this list (Listing \ref{lst:impl}).

\begin{lstlisting}[caption={SFQ component abstraction}, label={lst:impl}]
pub struct Wire {name: String}
enum Gates {
  And {name: String, a: String, b: String, clk: String, q: String},
  Split {name: String, a: String, q0: String, q1: String},
  ...
}
impl<const N: usize, ... > Circuit<N, M, L, O, P> {
  pub fn and(&mut self, a: Wire, b: Wire, clk: Wire, q: Option<&str>) -> Wire {
    let gate = Gates::And { ... };
    self.gates.push(gate);
    return Wire::new( ... );
  }
  ...
}
\end{lstlisting}

\begin{figure*}[ht]
    \centering
    \begin{minipage}[b]{0.24\textwidth}
        \centering
        \includegraphics[scale=0.1]{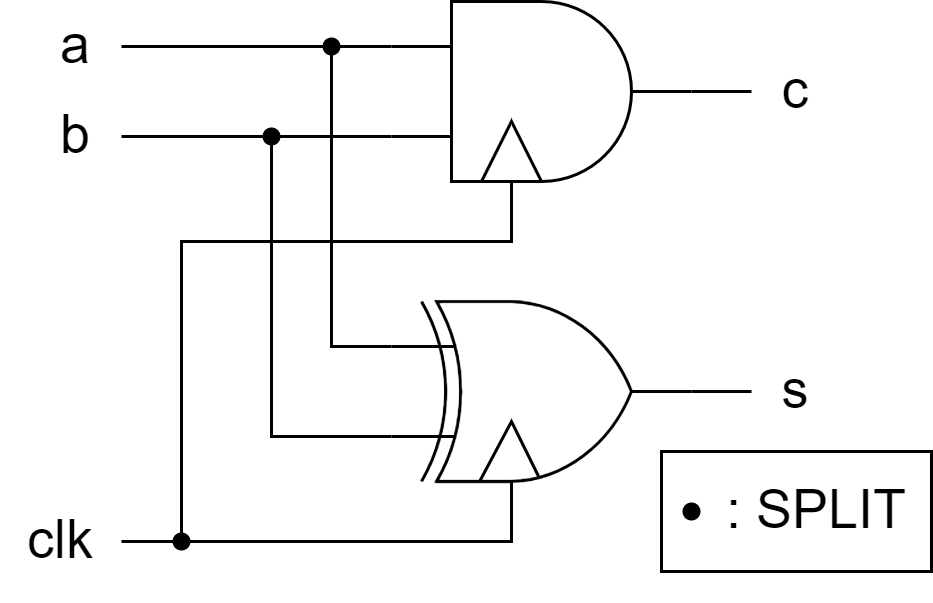}
        \caption{Half adder circuit}
        \label{halfadder}
    \end{minipage}
    \hfil
    \begin{minipage}[b]{0.41\textwidth}
        \centering
        \includegraphics[scale=0.1]{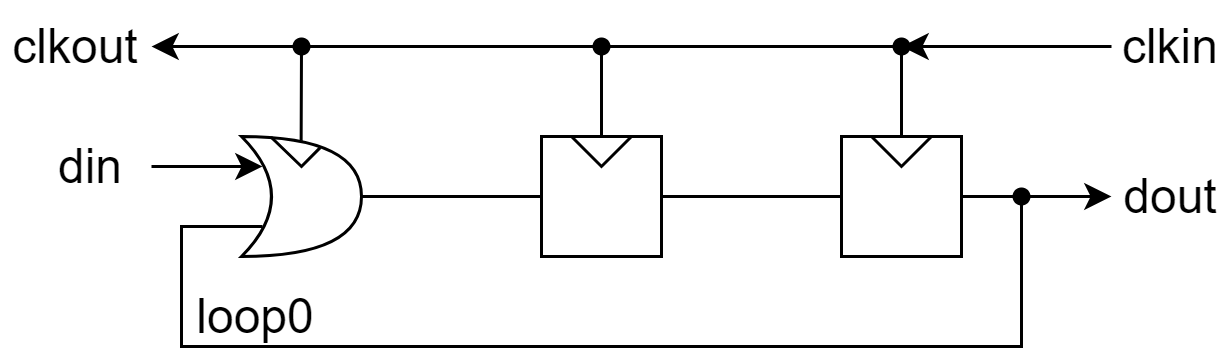}
        \caption{Counter-flow clocking circuit with a loop}
        \label{advanced}
    \end{minipage}
    \hfil
    \begin{minipage}[b]{0.28\textwidth}
        \centering
        \includegraphics[scale=0.11]{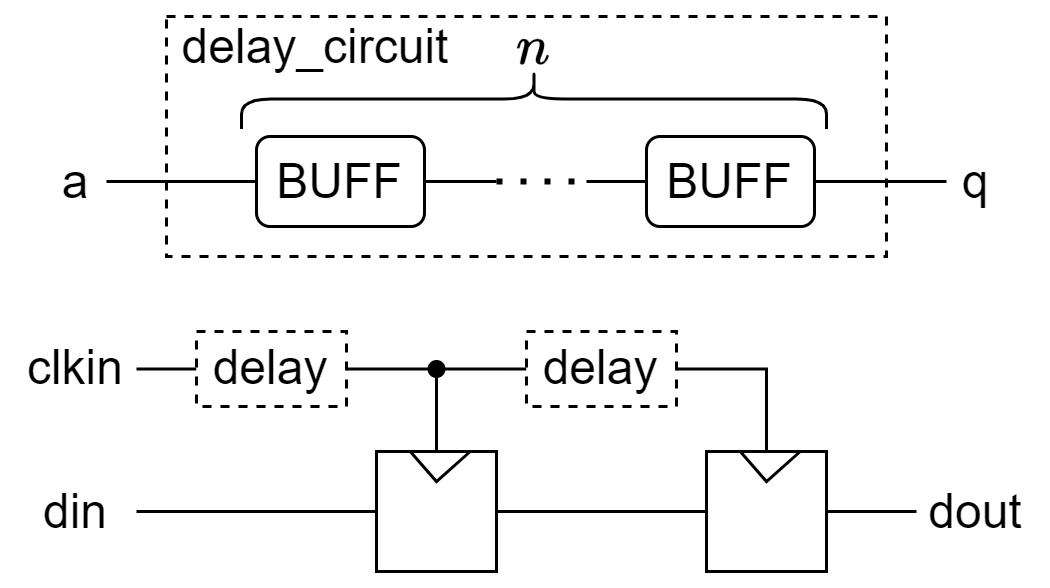}
        \caption{Parametrized delay circuit}
        \label{delay}
    \end{minipage}
\end{figure*}

\begin{lstlisting}[caption={DSL description for the circuit in \figurename~\ref{halfadder}}, label={lst:halfadder}]
let inputs = ["a", "b", "clk"];
let outputs = ["c", "s"];
let (mut circuit, [a, b, clk], [], []) =
    Circuit::create(inputs, outputs, [], [], [],
    "HalfAdder");
let [clk1, clk2] = circuit.split(clk, None, None);
let [a1, a2] = circuit.split(a, None, None);
let [b1, b2] = circuit.split(b, None, None);
let c = circuit.and(a1, b1, clk1, Some("c"));
let s = circuit.xor(a2, b2, clk2, Some("s"));
circuit.set_outputs([c, s]);
println!("{}", circuit.to_spice().join("\n"));
\end{lstlisting}

As a primary example, an SFQ-based half-adder circuit and its DSL description are shown in \figurename~\ref{halfadder} and Listing \ref{lst:halfadder}, respectively.

A \texttt{Circuit} instance, which corresponds a subcircuit in SPICE and a module in Verilog, is created by invoking the \texttt{Circuit::create()} function with its input and output ports specified. At the time, \texttt{Wire} instances for the inputs (\texttt{a}, \texttt{b}, and \texttt{clk} in Listing \ref{lst:halfadder}) are also created.

Logic gates are instantiated by invoking dedicated functions on the \texttt{Circuit}. Each function takes the \texttt{Wire}s for the gate's inputs and takes ownership of them. Consequently, the \texttt{Wire} cannot be used again. The gate function then returns a newly generated \texttt{Wire} for its output, which is used as an input for the subsequent gate.

A \texttt{Wire} can only be obtained from the circuit's input port or the gates' output, and can only be used exactly once. This mechanism ensures the \ioconsistency{} required in SFQ circuits.
The \texttt{Wire}s corresponding to the circuit's output port are passed to the \texttt{set\_output()} function, which consumes their ownership.

The \texttt{Circuit} has two functions \texttt{to\_spice()} and $\texttt{to\_verilog()}$, which generate netlists in SPICE or Verilog format as strings.
Conversion of the DSL code into netlists is completed by compiling and executing the code as a Rust code.

\subsection{Advanced Design}
In the RustSFQ, circuits are described from upstream to downstream, but certain types of circuits cannot be described in this manner. These include circuits with feedback loops and circuits employing counter-flow clocking, in which the clock signal travels in the opposite direction of the data flow.

To describe a circuit with loops, \texttt{Wire}s at the upper end of the loop are created along with the \texttt{Circuit}. The \texttt{Wire}s at the lower end are identified as the upper end using the \texttt{set\_loops()} function.

The \texttt{CounterWire} type is introduced to describe counter-flow clocking circuits. The DSL provides only BUFF and SPLIT gate functions for \texttt{CounterWire}, which are used in clock lines. In contrast to standard gate functions, these functions receive the \textit{output} \texttt{CounterWire} and return the \textit{input} \texttt{CounterWire}. Listing \ref{lst:advanced} and \figurename~\ref{advanced} illustrate these descriptions and show the resulting circuit structures.

\begin{lstlisting}[caption={DSL description for the circuit in \figurename~\ref{advanced}}, label={lst:advanced}]
let (mut c, [din], [loop0], [clkout]) = Circuit::create(
    ["din"], ["dout"], ["loop0"], ["clkout"],
    ["clkin"], "Advanced");

let (clk, clk0) = c.counter_split(clkout,None,None);
let d = c.or(din, loop0, clk0, None);
let (clk, clk0) = c.counter_split(clk, None, None);
let d = c.dff(d, clk0, None);
let (clkin, clk0) = c.counter_split(clk, 
    Some("clkin"), None);
let d = c.dff(d, clk0, None);
let [dout, loop0] = c.split(d, Some("dout"),
    Some("loop0"));
c.set_outputs([dout]);
c.set_loops([loop0]);
c.set_counter_inputs([clkin]);
\end{lstlisting}

The DSL enables circuits to be parametrized, which is difficult in netlists. For example, a delay circuit comprising a specified number of BUFF gates is described
in Listing \ref{lst:delay}. Here, Rust's ownership system is sufficiently strong to know that the \texttt{Wire} instance in variable \texttt{a} is consumed and created in every iteration; thus, the \ioconsistency{} is still ensured.

A \texttt{Circuit} can be reused multiple times in another \texttt{Circuit} using the \texttt{subcircuit()} function. The function takes the reused \texttt{Circuit} and \texttt{Wire}s for inputs, then returns \texttt{Wire}s for outputs. The number of inputs and outputs is statically checked with the circuit type, which includes the port counts.

\begin{lstlisting}[caption={DSL description for the circuit in \figurename~\ref{delay} ($n=5$)}, label={lst:delay}, float]
fn delay_circuit(n: u32) -> Circuit<1, 1, 0, 0, 0> {
  let name = format!("delay{}", n);
  let (mut c, [mut a], [], []) =
  Circuit::create(["a"], ["q"], [], [], [], &name);
  for i in 0..n {
    let name = if i == n - 1 { Some("q") } else { None };
    a = c.buff(a, name);
  }
  c.set_outputs([a]);
  return c;
}
fn main() {
  let delay5 = delay_circuit(5);
  let (mut c, [d, clk], [], []) = Circuit::create(
    ["din", "clk"], ["dout"], [], [], [], "main");
  let ([clk], []) = c.subcircuit(&delay5, [clk], [],[None],[]);
  let [clk, clk1] = c.split(clk, None, None);
  let d = c.dff(d, clk1, None);
  let ([clk], []) = c.subcircuit(&delay5, [clk], [],[None],[]);
  let d = c.dff(d, clk, Some("dout"));
  c.set_outputs([d]);
  println!("{}", delay5.to_spice().join("\n"));
  println!("{}", c.to_spice().join("\n"));
}
\end{lstlisting}

\subsection{Static Checking of Defective Circuits}
The pulse-driven nature of SFQ circuits requires that a pulse generated by one gate be received by exactly one gate. Circuits violating this restriction, such as those with more than one receiver or no receiver, are defective as SFQ circuits. RustDSL enables static detection of these problems, leveraging Rust's ownership system and static code analyses.

First, the circuit description in Listing \ref{lst:multireceivers} is defective because wire \texttt{a} is received by multiple gates, AND and XOR. When you try to compile this code, the Rust compiler gives an error about ownership. The variable \texttt{a} loses ownership at the first function call of AND, and its second use in the XOR function is prohibited.

Second, the circuit description in Listing \ref{lst:noreceiver} is also defective because no gates receive the wire \texttt{a3} after it is split. When you try to compile this code, the Rust compiler warns about an unused variable. 

On the contrary, a description without errors or warnings will guarantee that the circuit is valid as an SFQ circuit that satisfies \ioconsistency{}.

\begin{lstlisting}[caption={Defective description with multiple receivers}, label={lst:multireceivers}]
...
let c = circuit.and(a, b1, clk1, Some("c"));
let s = circuit.xor(a, b2, clk2, Some("s"));
circuit.set_outputs([c, s]);
\end{lstlisting}

\begin{lstlisting}[caption={Defective description with no receiver}, label={lst:noreceiver}]
...
let [a1, a2] = circuit.split(a, None, None);
let [a2, a3] = circuit.split(a2, None, None);
let c = circuit.and(a1, b1, clk1, Some("c"));
let s = circuit.xor(a2, b2, clk2, Some("s"));
circuit.set_outputs([c, s]);
\end{lstlisting}

\subsection{Automation of Net Name Assignment}
When writing a netlist, the assignment of net names and their error-free writing constitute a significant burden for circuit designers. Although it is easy to number each net in order, it increases the cost of circuit modification and debugging. On the other hand, naming everything with a meaningful name helps debugging but increases the cost of naming.
RustDSL provides a solution to this problem: when creating a \texttt{Wire}, i.e., when creating a \texttt{Circuit} or placing a gate, the wire can be optionally named. 

Signals for I/O or those important for debugging can be explicitly named. If the net name is omitted, the name is automatically assigned at compile time. Since the automatically generated names start with an underscore and are numbered consecutively, there is no conflict unless the user dares to break the naming rule.
Listing \ref{lst:netlist} shows the netlist output from the description of listing \ref{lst:halfadder}.

\begin{lstlisting}[caption={Netlist in SPICE format output from Listing \ref{lst:halfadder}}, label={lst:netlist}]
.subckt HalfAdder a b clk c s
XSPLIT1 clk _clk_0 _clk_1 THmitll_SPLIT
XSPLIT2 a _a_0 _a_1 THmitll_SPLIT
XSPLIT3 b _b_0 _b_1 THmitll_SPLIT
XAND4 _a_0 _b_0 _clk_0 c THmitll_AND2
XXOR5 _a_1 _b_1 _clk_1 s THmitll_XOR
.ends
\end{lstlisting}

\section{A Case Study on SFQ Reed--Solomon Encoder}
We demonstrated the practicality of the RustSFQ by developing a Reed--Solomon (RS) encoder circuit. It encoded an RS(12, 8) with a 4-bit width. The process was pipelined in four stages. It had four input buffers with eight depths, and its pipeline was filled by sequentially placing data from the buffers at every clock. It operated at a frequency of 10 GHz, as verified by digital and analog simulations.

\subsection{Reed--Solomon Code}
RS codes are error-correcting codes that are widely used in various applications, such as QR codes and wireless communications \cite{thach2014robust, daya2013fpga}.
Since SFQ circuits use magnetic flux to perform calculations, it has been reported that electromagnetic noise can cause errors in operation~\cite{yamanashi2007design}. RS codes and other error-correcting codes may be utilized to improve the noise tolerance of communication and memory in SFQ circuits.

RS codes vary depending on the bit width, message length, and error correction code length. An RS code is notated as RS$(n, k)$ if a message of $k$ words is encoded into $n$ words, with $n-k$ additional error-correcting words.
Let $n-k = 2t$, where $t$ represents the error correction capability. This implies that errors affecting up to $t$ words can be corrected.

Let $a$ be the bit width. Each word in an $a$-bit RS code is an element of the Galois field $\mathrm{GF}(2^a)$. 
The field $\mathrm{GF}(2^a)$ is a set consisting of $2^a$ elements, each of which can be represented in multiple forms: power, polynomial, and binary. 
For example, in $\mathrm{GF}(2^4)$ with generator polynomial $p(x) = 1 + x + x^4$, the elements can be expressed as listed in Table \ref{tab:galois}. 
Here, $\alpha$ satisfies $p(\alpha) = 0$ and is referred to as the primitive element.

In the polynomial representation, the coefficients are elements of $\mathrm{GF}(2)$, which is the set $\{0, 1\}$. 
The binary representation corresponds to the sequence of these coefficients. 
Addition in $\mathrm{GF}(2^a)$ is performed as an XOR operation on the binary representations, 
whereas multiplication is performed as the product of the power representations. 
For multiplication, the exponents are considered modulo $2^a - 1$.

In the RS$(n, k)$ code, a $k$-word message $(m_0, m_1, \dots, m_{k-1})$, 
where $m_i \in \mathrm{GF}(2^a)$ for all $i$, can be represented as a $(k-1)$th-degree polynomial:
\[
m(x) = m_0 + m_1x + \dots + m_{k-1}x^{k-1} \in \mathrm{GF}(2^a)[x].
\]

Here, a generator polynomial $g(x) \in \mathrm{GF}(2^a)[x]$ of degree $n-k$ is used, 
and the parity polynomial $p(x) \in \mathrm{GF}(2^a)[x]$ \hspace{1cm} of degree $n-k-1$ is computed as the remainder when \hspace{1cm} $m(x) \cdot x^{n-k}$ is divided by $g(x)$.

Let $p(x)$ be expressed as
\[
p(x) = p_0 + p_1x + \dots + p_{n-k-1}x^{n-k-1},
\]
where $(p_0, p_1, \dots, p_{n-k-1})$ represents the error-correcting code. 
The codeword is then formed by concatenating the message and the error-correcting code.

For example, consider a 4-bit RS$(12, 8)$ code with a generator polynomial 
\[
g(x) = \alpha^{10} + \alpha^3 x + \alpha^6 x^2 + \alpha^{13} x^3 + x^4.
\]
Suppose the message is $(0010, 0101, 0011, 0101, 0110, 0100,\\ 0001, 0101)$. 
This corresponds to the following polynomial:
\[
m(x) = \alpha^2 + \alpha^9 x + \alpha^6 x^2 + \alpha^9 x^3 
+ \alpha^5 x^4 + \alpha^1 x^5 + \alpha^3 x^6 + \alpha^9 x^7.
\]
The remainder of dividing $m(x) \cdot x^4$ by $g(x)$ is
\[
p(x) = 0 + \alpha^2 x + \alpha x^2 + \alpha^{11} x^3.
\]
Thus, the error-correcting code is calculated as $(0000, 0010, 0100, 0111)$.

The error detection and correction processes of the RS codes are omitted in this paper.

\begin{table}[t]
    \centering
    \begin{tabular}{|c|c|c|}
        \hline
        Power & Polynomial & Binary \\
        Representation & Representation & Representation \\
        \hline
        0 & $0 \phantom{{}+ \alpha + \alpha^2 + \alpha^3}$ & 0000 \\
        1 & $1 \phantom{{}+ \alpha + \alpha^2 + \alpha^3}$ & 1000 \\
        $\alpha$ & $\phantom{1+{}}\alpha\phantom{{}+\alpha^2+\alpha^3}$ & 0100 \\
        $\alpha^2$ & $\phantom{1+\alpha+{}}\alpha^2\phantom{{}+\alpha^3}$ & 0010 \\
        $\alpha^3$ & $\phantom{1+\alpha+\alpha^2+{}}\alpha^3$ & 0001 \\
        $\alpha^4$ & $1+\alpha\phantom{{}+\alpha^2+\alpha^3}$ & 1100 \\
        $\alpha^5$ & $\phantom{1+{}}\alpha+\alpha^2\phantom{{}+\alpha^3}$ & 0110 \\
        $\alpha^6$ & $\phantom{1+\alpha+{}}\alpha^2+\alpha^3$ & 0011 \\
        $\alpha^7$ & $1+\alpha\phantom{{}+\alpha^2}+\alpha^3$ & 1101 \\
        $\alpha^8$ & $1\phantom{{}+\alpha}+\alpha^2\phantom{{}+\alpha^3}$ & 1010 \\
        $\alpha^9$ & $\phantom{1+{}}\alpha\phantom{{}+\alpha^2}+\alpha^3$ & 0101 \\
        $\alpha^{10}$ & $1+\alpha+\alpha^2\phantom{{}+\alpha^3}$ & 1110 \\
        $\alpha^{11}$ & $\phantom{1+{}}\alpha+\alpha^2+\alpha^3$ & 0111 \\
        $\alpha^{12}$ & $1+\alpha+\alpha^2+\alpha^3$ & 1111 \\
        $\alpha^{13}$ & $1\phantom{{}+\alpha}+\alpha^2+\alpha^3$ & 1011 \\
        $\alpha^{14}$ & $1\phantom{{}+\alpha+\alpha^2}+\alpha^3$ & 1001 \\
        \hline
    \end{tabular}
    \caption{Representations of elements in GF($2^4$) (adapted from \cite{sylvester2001reed})}
    \label{tab:galois}
\end{table}

\subsection{Existing CMOS Implementation of RS Encoder}
\begin{figure}[t]
    \centering
    \includegraphics[width=\linewidth]{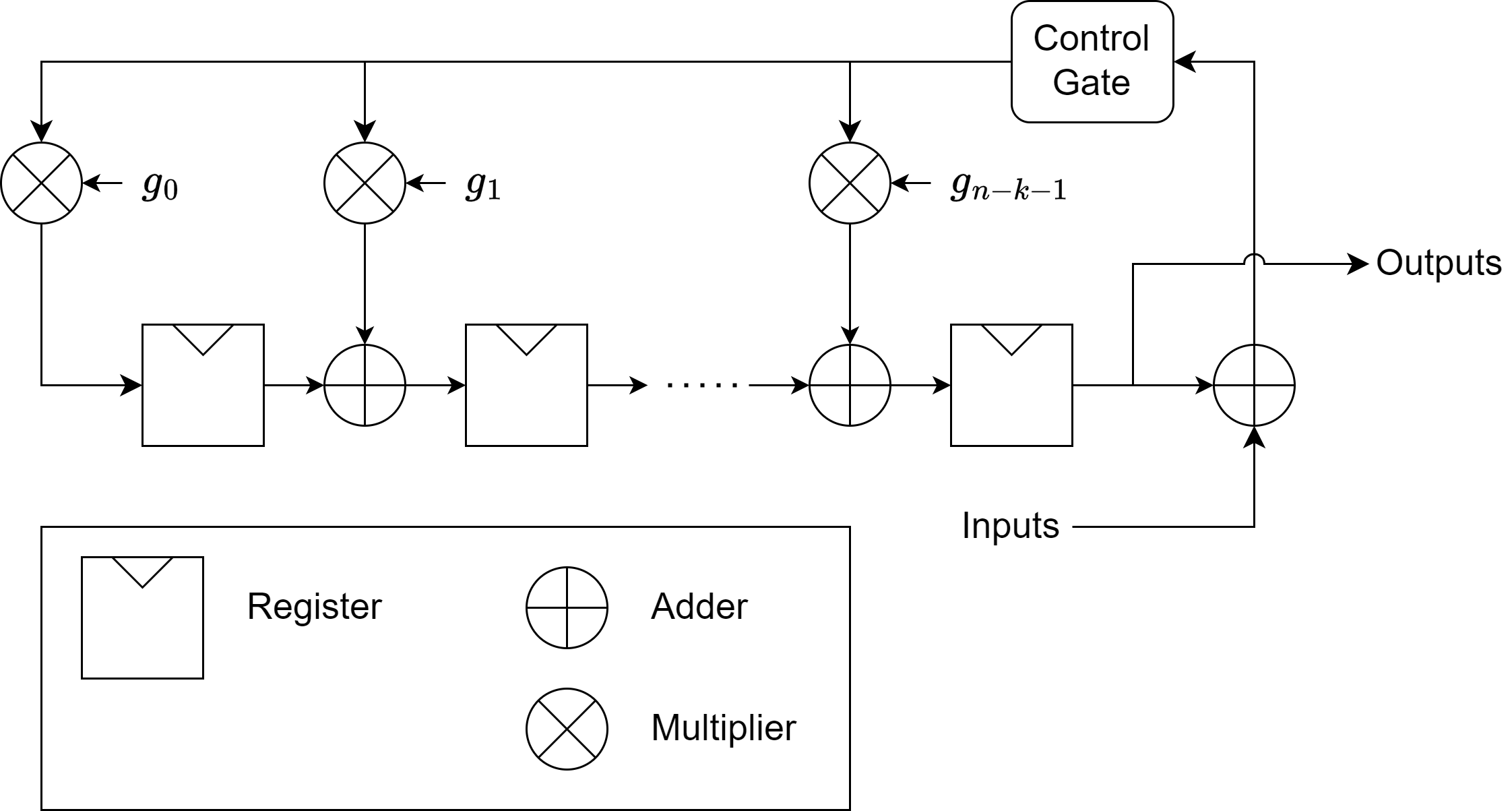}
    \caption{A CMOS implementation of RS encoder}
    \label{fig:fig12}
\end{figure}

A typical hardware RS encoder \cite{sylvester2001reed} can be implemented as shown in \figurename~\ref{fig:fig12}.
The registers are initialized to zero. The message is entered sequentially from the highest to lowest coefficient. Multipliers and adders are combinational circuits that perform computations in the Galois field. The control gate manages the feedback signals and initially allows them to pass. After $k$~words are input, the registers contain error-correcting words. At this point, the gate is shut down, which means that zeros are input into the multipliers. The results can then be obtained in the subsequent $n-k$ cycles.

\subsection{Designed SFQ Implementation of RS Encoder}
Owing to the differences in the behavior of logic gates, circuits designed for CMOS circuits cannot be directly applied to SFQ circuits. we developed an RS encoder that operates efficiently in SFQ circuits based on the circuit shown in \figurename~\ref{fig:fig12}.

\begin{figure*}[ht]
    \centering
    \includegraphics[width=\linewidth]{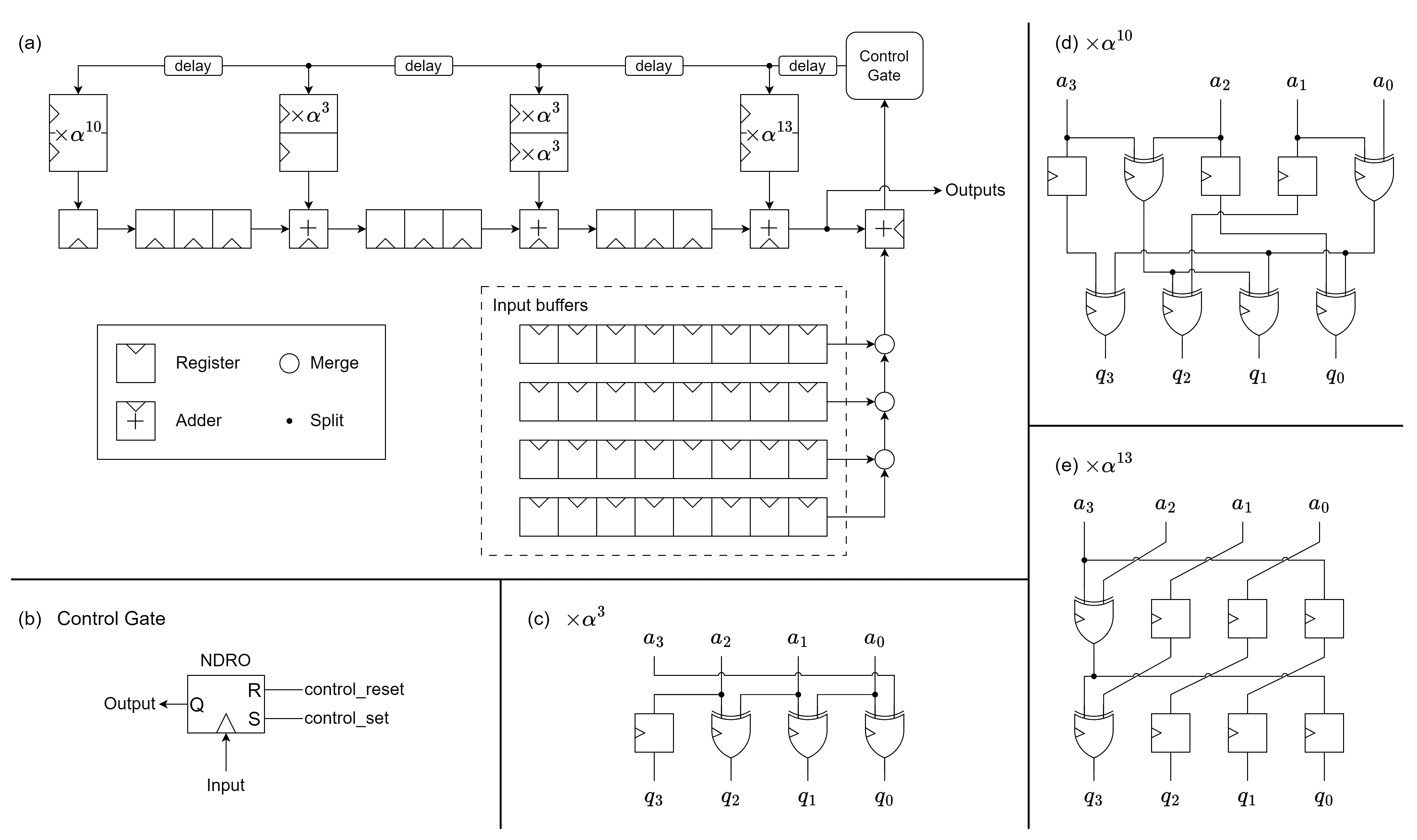}
    \caption{(a) Overall designed SFQ Reed--Solomon encoder, (b) Clockless gating circuit with an NDRO in Control Gate, \\
    (c), (d), (e) Galois field constant multipliers}
    \label{fig:sfqrs}
\end{figure*}

\figurename~\ref{fig:sfqrs} shows the designed RS encoder for RS(12,8) with a 4-bit width, which, to the best of our knowledge, is the first implementation of an RS encoder in SFQ circuits.

\subsubsection{Constant Multipliers}

The RS encoder utilizes Galois field multipliers, but only specific constant multiplications are necessary, rather than general multiplication. The RS(12, 8) encoder requires multiplications by four constants: $\alpha^3$, $\alpha^6$, $\alpha^{10}$, and $\alpha^{13}$. Each of these constant multipliers was designed as a separate circuit.

\figurename~\ref{fig:sfqrs}. (c), (d), and (e) show the constant multipliers that consist of XOR gates and DFFs for timing adjustment. These circuits were manually optimized to reduce the number of logic-gate stages. Although the $\alpha^3$ multiplier can complete its computation in a single stage, an additional DFF stage is added to match its stage count with that of the other multipliers. The $\alpha^6$ multiplier comprises two $\alpha^3$ multipliers.

\subsubsection{Pulse Gating}
In CMOS circuits, the gating of a data signal with a control signal can be easily realized using only an AND gate. However, in SFQ circuits, an AND gate requires a clock, which is unfavorable in some cases. 

A clockless gating circuit is realized using a non-destructive readout (NDRO) gate. The internal states of an NDRO gate are similar to those of an RS flip-flop. Its state is updated whenever a set or reset input occurs. When a clock is given, it outputs a pulse if its internal state is ``1''. An NDRO gate differs from other SFQ gates in that its internal state is not changed by clocks.

To use an NDRO gate as a gating circuit, the controlled data line is connected to its \textit{clock} input (\figurename~\ref{fig:sfqrs}. (b)). When the gate's internal state is ``1'', it outputs a pulse every time a data pulse comes, which looks like the gate passes the input pulses. When its internal state is ``0'', it outputs no pulse even when a data pulse comes.

\subsubsection{Pipelining}
After inputting a word into the encoder, the output value of the adder next to the $\alpha^{13}$ multiplier is added to the following input value to be feedbacked. Because two adders requires one cycle each, the constant multiplier requires two cycles, and the control gate requires no cycle, a set of data must be input every four cycles.

However, because independent data can be input without waiting, pipeline execution can be performed using four input sets. The implemented RS encoder has four 8-depth shift registers as input buffers, sequentially feeding the input value into the calculation circuit to maximize circuit utilization.

\subsubsection{Clocking}
In SFQ circuits that include feedback loops, it is known that higher operating frequency can be achieved by using counter-flow clocking, in which the clock delay can hide the data delay in the loop section \cite{ishida2020supernpu}.
Therefore, we employed counter-flow clocking in the RS encoder.
The clock takes 500 ps from the farthest downstream to the farthest upstream, but the encoder operates with a clock period of 100 ps. Appropriate delays must be inserted into the feedback section to ensure the correct operation. These delays must not be too small or too large because the data and clocks must arrive in the correct order.

\section{Simulation Results and Evaluation}
The designed RS encoder circuit's operation was verified through digital simulation using Icarus Verilog and analog simulation using JoSIM. The input consisted of the following four messages:
\begin{enumerate}
  \item (0100, 1110, 0111, 1111, 0100, 0010, 0001, 1100)
  \item (0001, 0010, 0011, 0100, 0101, 0110, 0111, 1000)
  \item (1000, 1000, 1000, 1000, 1000, 1000, 1000, 1000)
  \item (0010, 0101, 0011, 0101, 0110, 0100, 0001, 0101)
\end{enumerate}

The expected error-correcting codes for these messages are as follows:
\begin{enumerate}
  \item (1000, 0011, 0011, 1101)
  \item (1010, 1010, 0001, 0001)
  \item (0101, 0001, 0110, 0010)
  \item (0000, 0010, 0100, 0111)
\end{enumerate}

\begin{figure}[ht]
    \centering
    \includegraphics[width=\linewidth]{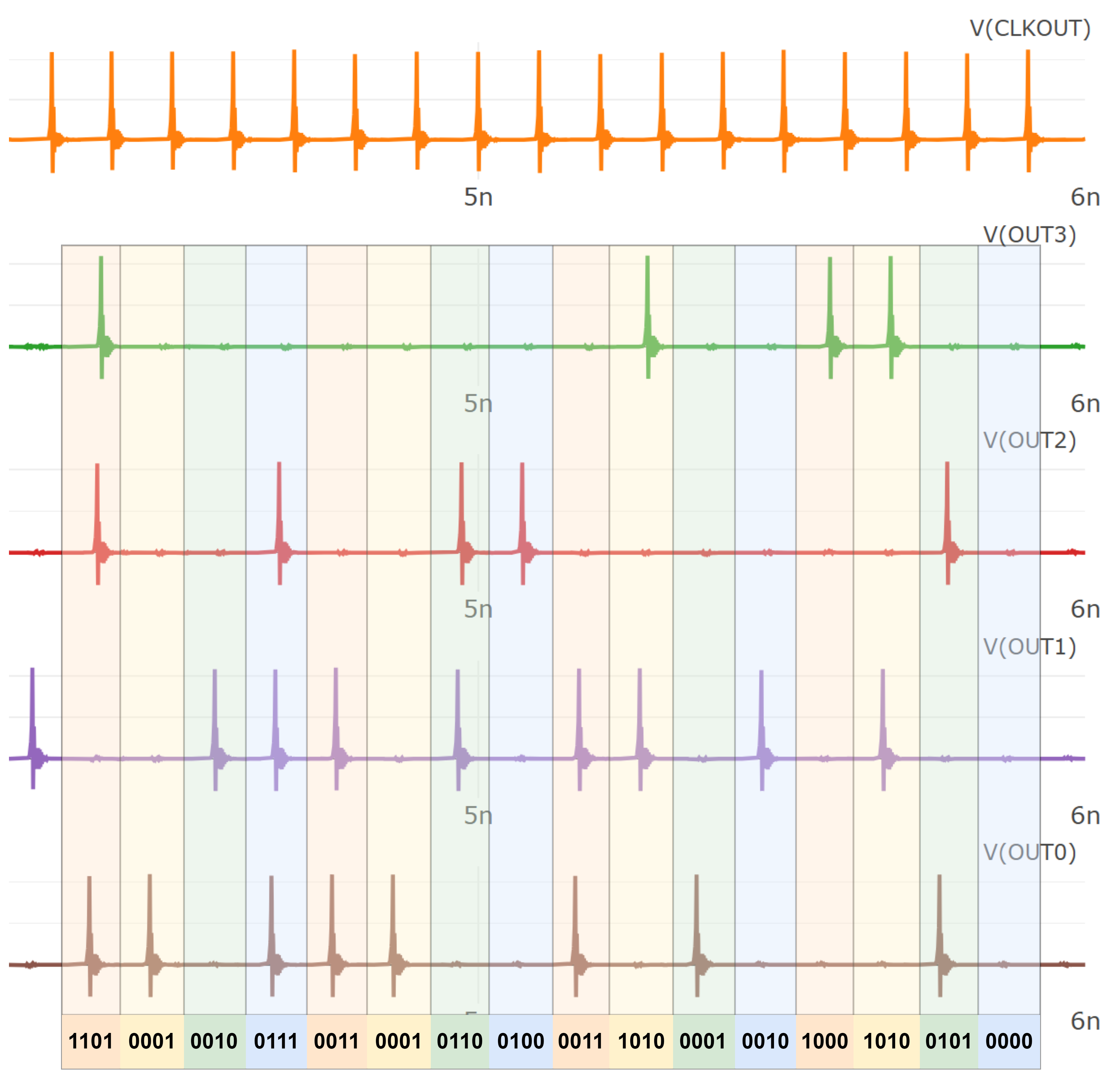}
    \caption{Result of analog simulation of the RS encoder.}
    \label{fig:fig16}
    \footnotesize{The four colors correspond to each set of outputs. The results are output in order from the highest-order coefficient.}
\end{figure}

\figurename~\ref{fig:fig16} shows some of the results of the analog simulation. This illustrates that the four sets of results were correctly output in sequence from the highest-order value.
Excluding the time required to store data into the input buffer, the computation takes 48 cycles. Since the clock period was 100 ps, the calculation completed in 4.8 ns.

The digital simulation also confirmed that the circuit operated correctly.

The circuit of the RS encoder was described using the RustSFQ. The DSL code consisted of two files with a total file size of 24,885 bytes. After compiling the DSL code, the size of the resulting SPICE file was 29,837 bytes and that of the Verilog file was 49,256 bytes (\tablename~\ref{tab:filesize}).
Compared with the combined size of the output files, the DSL source code size was less than one-third, indicating a reduction in the amount of code required for the description.

Furthermore, one of the burdens of writing netlists, assigning net names, was notably eased. This circuit contained 796 net names, of which only 218 were specified explicitly (\tablename~\ref{tab:assignment}). Over 70 \% of the net names were automatically assigned during compilation, relieving designers of the need to assign them manually.

Another advantage of using the DSL is its ability to detect errors earlier. In addition to syntax and spelling mistakes, any input--output inconsistencies specific to SFQ circuits can be identified during the compilation stage. Moreover, in most cases, these errors can be identified in real time without compilation thanks to the Rust language server integrated into the editors.

\begin{table}[h]
    \centering
    \begin{tabular}{c|c}
        Code & File Size (bytes) \\
        \hline
        RustSFQ & 24,885 \\
        \hline
        SPICE & 29,837 \\
        Verilog & 49,256 \\
    \end{tabular}
    \caption{File sizes of codes to describe the RS encoder}
    \label{tab:filesize}
\end{table}

\begin{table}[h]
    \centering
    \begin{tabular}{c|c c}
        Name Assignment & Count \\
        \hline
        Manual & 218 (27.4 \%)\\
        Automatic & 578 (72.6 \%)\\
    \end{tabular}
    \caption{Net name assignments of the RS encoder}
    \label{tab:assignment}
\end{table}

\section{Conclusion}
RustSFQ, a DSL for SFQ circuit design that is embedded in Rust, was proposed. 
It enables a higher level of abstraction while maintaining fine-grained descriptions as conventional netlists. It can fill the lack of tools for efficient SFQ circuit design.
The key feature of RustSFQ is that it ensures the \ioconsistency{} by leveraging Rust's ownership system. Other features, such as automatic assignment of net names and compatibility for both analog and digital simulations, contribute to effective SFQ circuit design.

\section*{Acknowledgements}
This work is supported in part by JSPS KAKENHI Grant Number 23H00467.

\bibliography{myref}

\end{document}